\documentclass{moriond}

\usepackage[english]{babel}
\usepackage{amsmath}
\usepackage{cleveref}
\usepackage{graphicx,xcolor}
\usepackage[super,sort&compress]{natbib}

\def\be{\begin{equation}}
\def\ee{\end{equation}}
\def\bea{\begin{eqnarray}}
\def\eea{\end{eqnarray}}

\newcommand{\BR}{\text{BR}}

\newcommand{\Photo}{\includegraphics[height=35mm]{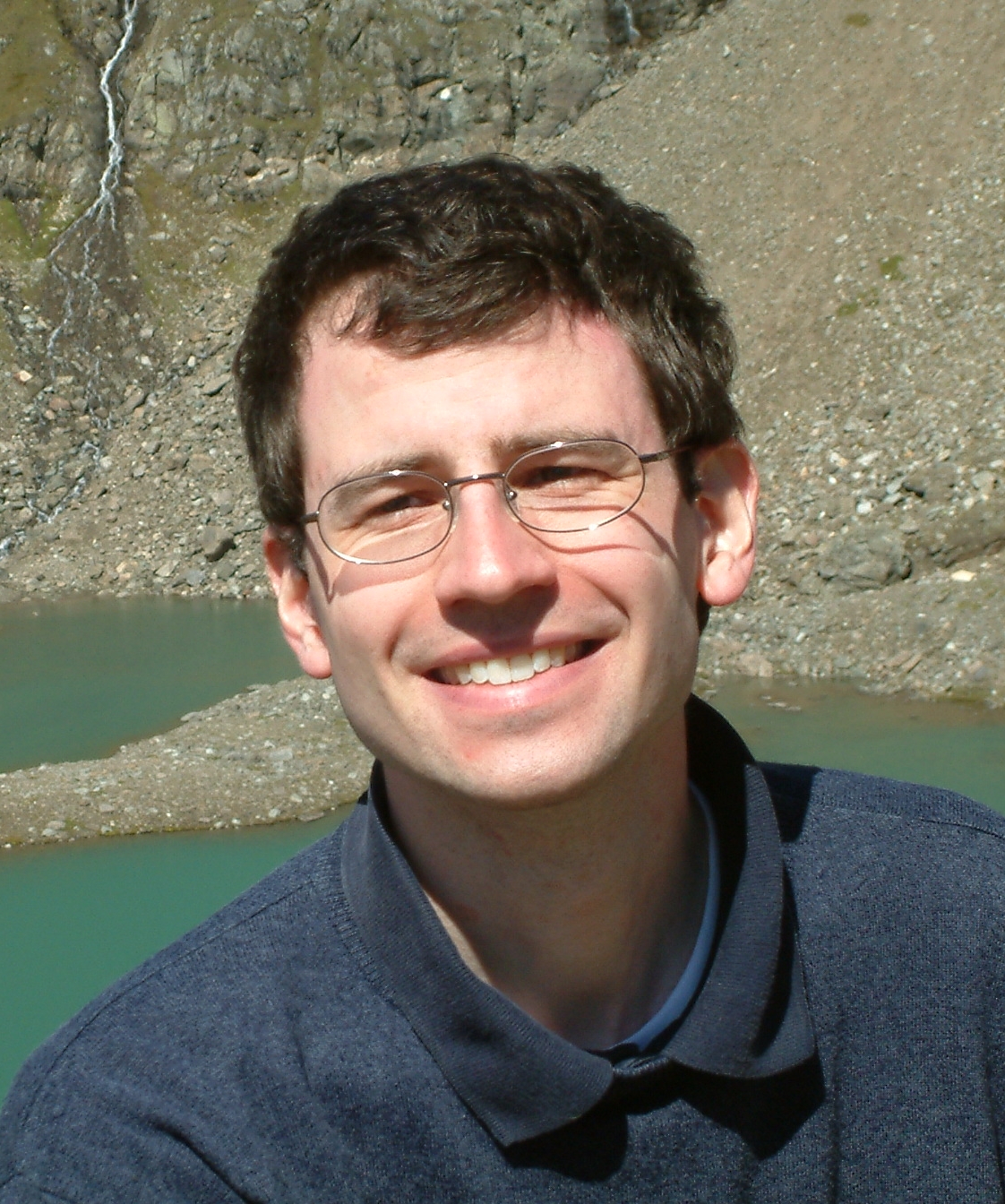}}

\begin{document}

\vspace*{4cm}
\title{FLAVOR VIOLATION IN THE SCALAR SECTOR}

\author{J. KOPP}

\address{PRISMA Cluster of Excellence \& Mainz Institute for Theoretical Physics, \\
         Johannes Gutenberg University, Staudingerweg 7, 55128 Mainz, Germany}

\maketitle

\abstracts{
  In many extensions of the Standard Model, the alignment in flavor space of
  the fermion mass matrices and the Yukawa coupling matrices can be broken.
  The physical scalar boson $h(125)$ could then have flavor changing couplings.
  In this talk, we summarize constraints on such couplings from rare decay
  searches, and we investigate current and future detection prospects at the LHC.
  We emphasize the importance of several yet unexplored final states:
  (i) anomalous single top + $h$ production in $ p p \to t h$, arising
  from $tuh$ couplings (but not from the more widely studied $tch$ couplings);
  (ii) $p p \to t + (H^0 \to h h)$ through $tuh$ couplings in the context of a 
  Two Higgs Doublet Model (2HDM), perhaps the simplest model with flavor violation
  in the scalar sector; (iii) $p p \to H^0 \to \tau\mu$ in the 2HDM context.
  For all of these processes, we perform a detailed phenomenological studies.
  Finally, we comment on the possibility of flavor violation combined with
  CP violation, which may be interesting if the current CMS hint for $h \to \tau\mu$
  gets corroborated.
}

\section{Introduction}

The 125~GeV scalar boson discovered by ATLAS and CMS in 2012, while being the
last missing piece of the Standard Model (SM), will hopefully also contribute
to its demise by deviating from the expected behavior.  One such deviation
could be a flavor off-diagonal coupling to
 fermions,~\cite{Bjorken:1977vt,McWilliams:1980kj,Blankenburg:2012ex,Harnik:2012pb}
which are forbidden in the SM, but are quite naturally expected in
many of its extensions.
At the effective field theory level, flavor violating scalar couplings
have the generic structure (illustrated here for the charged leptons)
\begin{align}
  \mathcal{L} \supset
    -\frac{\eta_{ij}}{\Lambda^2} \bar{L}^i \tilde{H} e_R^j (H^\dag H)
  \quad\to\quad
    -y_{ij} \bar{e}_L^i e_R^j h + \cdots \,.
  \label{eq:L-eft}
\end{align}
Here, $\Lambda$ is the cutoff scale of the effective theory, $L^i$ are the
three left-handed SM lepton doublets, $e_R^j$ are the three right-handed
charged lepton singlet, $H$ is the SM scalar doublet, $\tilde{H} \equiv i
\sigma^2 H^\dag$, and $h$ is the physical 125~GeV scalar boson. Flavor violating couplings in
the quark sector are completely analogous. Possible ultraviolet completions of the
Lagrangian in \cref{eq:L-eft} exist for instance in Randall--Sundrum
models~\cite{Blanke:2008zb,Casagrande:2008hr,Azatov:2009na}, supersymmetric
models~\cite{DiazCruz:1999xe,deLima:2015pqa,Arhrib:2012mg,Aloni:2015wvn}, models
aiming to explain the flavor structure of the Standard
Model~\cite{Dery:2013rta}, leptoquark models~\cite{Cheung:2015yga}, and in
particular in Two Higgs Doublet Models
(2HDMs).~\cite{Bjorken:1977vt,DiazCruz:1999xe,Han:2000jz,Crivellin:2015mga,Omura:2015nja,Dorsner:2015mja,Crivellin:2015hha,Botella:2015hoa,Arhrib:2015maa,Benbrik:2015evd}

Phenomenologically, the operators in \cref{eq:L-eft} can lead to a bonanza
of flavor physics observables (rare decays, anomalous electric and magnetic
dipole moments, meson oscillations, \ldots), flavor changing scalar decays like
$h \to \tau\mu$, $h \to \tau e$, anomalous top quark decays $t \to h q$, and
anomalous single top + $h$ production via $u g \to t h$.  In UV completions
of \cref{eq:L-eft}, many more processes can be important. For instance in
the context of 2HDMs, flavor violating couplings of the heavy scalar bosons
are expected to be much larger than those of the lightest scalar, so that
the reactions $u g \to t H^0$ and $p p \to H^0 \to
\tau \mu$ can have sizeable rates.  In the following, we discuss the
aforementioned processes in more detail.

\section{Low Energy Constraints and Current LHC limits on FCNC in the Scalar Sector}

\begin{figure}
  \begin{center}
    \begin{tabular}{cc}
      \includegraphics[width=0.48\textwidth]{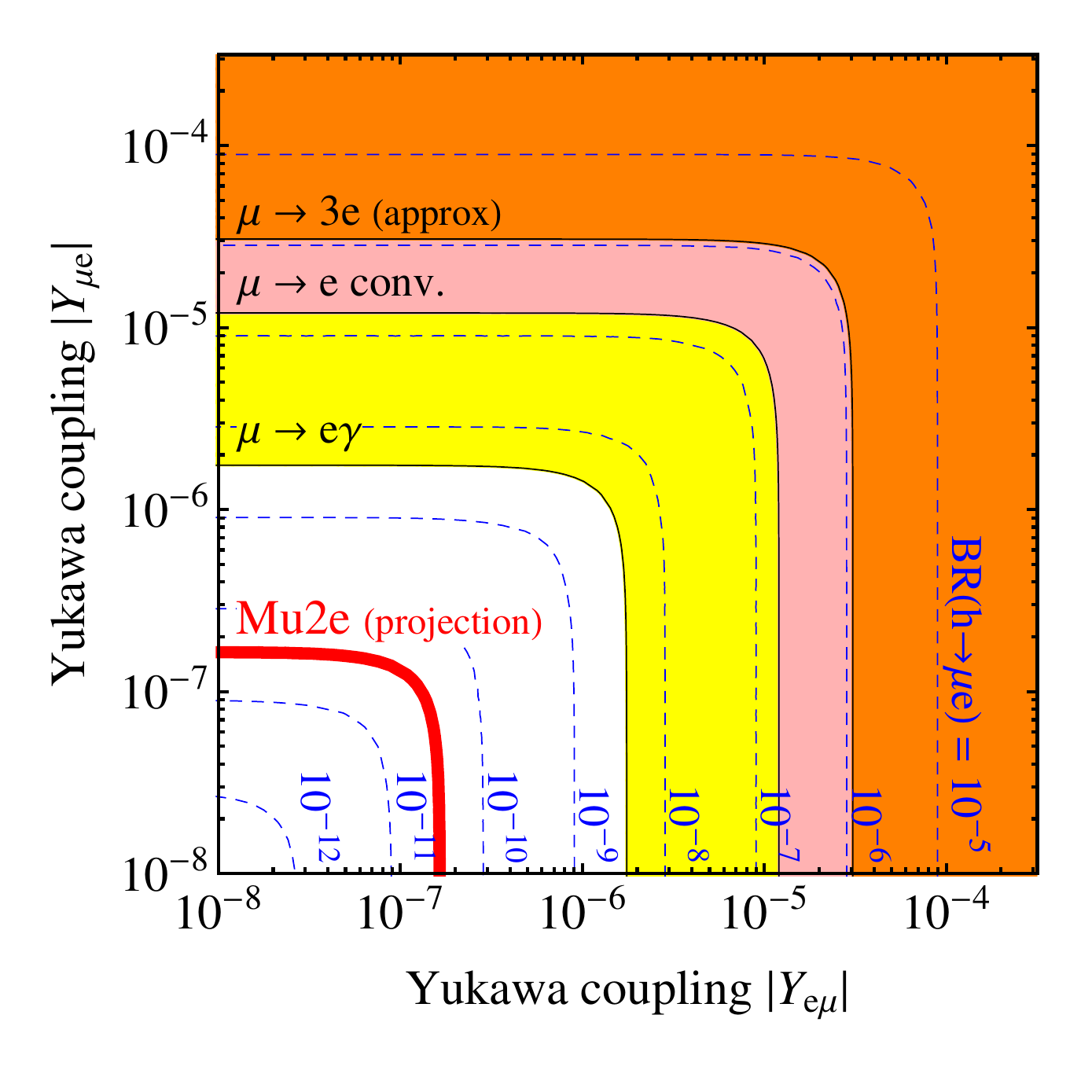}
      \raisebox{0.3cm}{\includegraphics[width=0.48\textwidth]{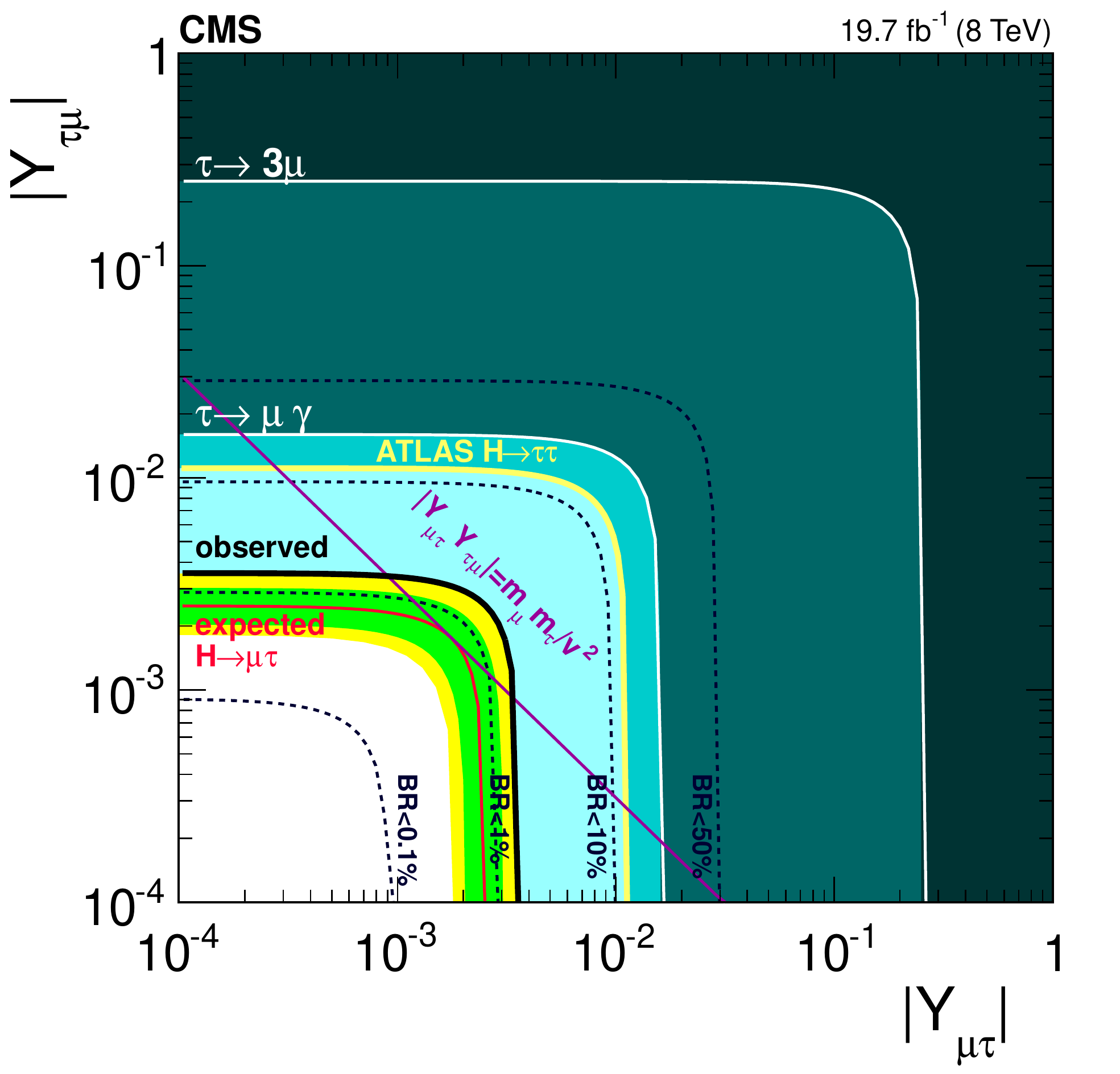}}
    \end{tabular}
  \end{center}
  \vspace{-0.3cm}
  \caption{Constraints on flavor changing couplings of the $h$ boson to leptons,
    expressed in terms of the Yukawa couplings appearing in \cref{eq:L-eft}.
    In the $\mu$--$e$ sector (left panel~\protect\cite{Harnik:2012pb}),
    low energy limits are many orders of
    magnitude stronger than LHC constraints (not shown here), while in the $\mu$--$\tau$
    sector (right panel~\protect\cite{Khachatryan:2015kon}), the LHC dominates.
    Note the small ($2.3\sigma$) excess in the CMS data on $h \to \mu\tau$.}
  \label{fig:lepton-constraints}
\end{figure}

Flavor changing neutral current (FCNC) couplings of the $h$ boson to leptons are most strongly
constrained by $\mu$--$e$ conversion in nuclei, by flavor changing decays
of a heavy lepton to three lighter leptons (e.g.\ $\tau \to 3\mu$), and by the radiative
decays $\mu \to e\gamma$, $\tau \to e\gamma$, and $\tau \to \mu\gamma$.
Currently, radiative decay limits have a slight edge over other constraints,
but with future experiments like Mu2e at Fermilab or Mu3e at PSI this will
change.  We show the current constraints on flavor changing neutral current (FCNC)
couplings of the
form $h \mu e$ in \cref{fig:lepton-constraints} (a) and on couplings
in the $\mu$--$\tau$ sector in \cref{fig:lepton-constraints} (b).
(Constraints on the $e$--$\tau$ sector are very similar to those on the
$\mu$--$\tau$ sector~\cite{Harnik:2012pb}.)
We see that constraints in the $\mu$--$e$ sector are so tight that the
associated LHC process $h \to \mu e$ is far beyond the experimental reach.
On the other hand, $\BR(h \to \tau\mu)$ or $\BR(h \to \tau e)$ can
be at the per cent level, well within the capabilities of ATLAS and CMS.
In fact, CMS has observed a small ($2.3\sigma$) excess in
$h \to \tau \mu$~\cite{Khachatryan:2015kon}.
(Note that \emph{either} $\BR(h \to \tau\mu)$ \emph{or}
$\BR(h \to \tau e)$ can be large, but not both. If both were sizeable,
the tightly constrained decay $\mu \to e \gamma$ would be induced at 1-loop.)

In the quark sector, FCNC processes are tightly constrained by low energy measurements
as well.  For instance, couplings of the form $h q q'$ (where $q,q' = u, d, s, c, b$)
contribute at tree level to neutral meson mixing.  The resulting
constraints~\cite{Harnik:2012pb} imply that the only flavor violating $h$ couplings
that could
be large enough to be observable at the LHC are those involving the top quark,
$tuh$ and $tch$. (Once again, only one of these couplings can be large, but not
both, to avoid large contributions to $D^0$--$\bar{D}^0$ meson mixing through box diagrams.)
The current experimental limits on FCNC top--$h$ couplings are
\begin{align}
  \BR(t \to c h) < 0.0046,
  \qquad\qquad
  \BR(t \to u h) < 0.0045
\end{align}
from ATLAS~\cite{Aad:2015pja} and $\BR(t \to u h) < 0.0047$,
$\BR(t \to c h) < 0.0042$ from CMS~\cite{CMS:2015xqa}.
Both limits are based on combinations of several final states, including in
particular multi-lepton and lepton + di-photon signatures.

\section{New Probes of FCNC Couplings to Quarks}

In the following, we outline several possible routes towards a further improvement
of the constraints on $tuh$ and $tch$ couplings.

\subsection{$t \to h q$ and $p p \to t h$}

First, we observe that $tuh$ couplings induce not only the widely studied decay
$t \to h u$, but also the process $u g \to t h$ (anomalous single top + $h$
production), which is not included in most searches for FCNC top
couplings~\cite{Greljo:2014dka}.  Therefore, the limits on $tuh$ couplings
reported by these searches are on the conservative side.  For the multi-lepton
and lepton + di-photon final states, it was shown~\cite{Greljo:2014dka} that
inclusion of anomalous single top + $h$ production could lead to an increase in
sensitivity by a factor $\sim 1.5$.
Note that this applies only to $tuh$ couplings, but not to $tch$ couplings.
The reason is that the process $c g \to t h$ is suppressed by the small charm quark
parton distribution function (PDF) and therefore
negligible.

\begin{figure}
  \begin{center}
    \includegraphics[width=0.45\textwidth]{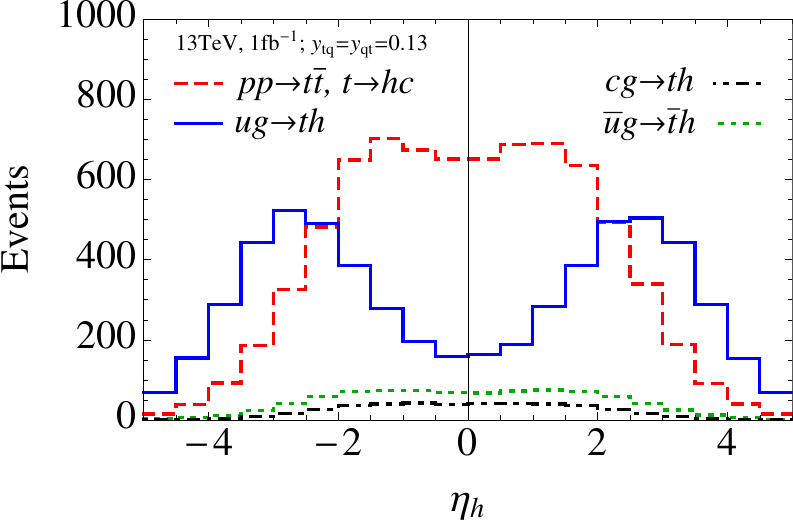}
  \end{center}
  \vspace{-0.3cm}
  \caption{The parton level pseudorapidity ($\eta_h$) distributions of the $h$
    boson in $p p \to (t \to W b) + (t \to h q)$ and in $u g \to t h$.~\protect\cite{Greljo:2014dka}
    Note the preference for forward emission of the $h$ boson in the latter
    process (see text for details).}
  \label{fig:higgs-eta}
\end{figure}

This observation suggests a possible way of distinguishing $tuh$ and $tch$
couplings in case of a discovery. Namely, even though the final states for $p p
\to (t \to W b) + (t \to h q)$ and $u g \to t + h$ are almost identical, there are
differences: in particular, the pseudorapidity ($\eta_h$) distribution of the $h$
boson and the sum of lepton charges in multi-lepton final states can be used
as discriminants~\cite{Khatibi:2014via,Greljo:2014dka}.  In $u g \to t
h$, the $h$ boson is preferentially emitted in the forward or backward
direction, compared to a more central distribution in $p p \to (t \to W b) + (t \to h
q)$, see \cref{fig:higgs-eta}.
The reason is that the chiral structure of the $tuh$ coupling implies a
chirality flip on the quark line. This is easily possible
only if the $t$ quark is emitted opposite to the direction of travel of the
initial $u$ quark. The $h$ boson must then travel preferentially in the same direction as
the $u$ quark.  The forward boost of the $h$ boson is further enhanced by the
fact that the center-of-mass frame of the process tends to be boosted in the
direction of the (valence) $u$ quark. The discrimination power of the sum of
lepton charges can be understood by noting that $u g \to t h$ events are much
more frequent than $\bar{u} g \to \bar{t} + h$ events because of PDF
suppression.  Quantitatively, one finds that, for a signal that is discovered
at the $5\sigma$ level, a $2\sigma$ discrimination between the $tuh$ and $tch$
hypotheses is possible~\cite{Greljo:2014dka}.  This estimate is for a multi-lepton
analysis and takes into account backgrounds, combinatorial uncertainties
and detector effects.

Going beyond the traditional multi-lepton and lepton + di-photon searches, a
further improvement of the sensitivity to $tuh$ and $tch$ couplings by an
$\mathcal{O}(1)$ factor is possible by including so-far unexplored final
states~\cite{Greljo:2014dka}, for instance the fully hadronic processes $p p
\to (t \to W b) + (t \to h q) \to \text{hadrons}$ and $p p \to (t \to W b) + h
\to \text{hadrons}$. These decays can be successfully reconstructed and
exploited using boosted object
taggers~\cite{Plehn:2009rk,Plehn:2010st,Butterworth:2008iy,Dokshitzer:1997in,Cacciari:2011ma}.

\subsection{$p p \to t h h$}

In specific models, additional search channels for flavor changing $h$
couplings lend themselves to exploitation.  Consider in this context a general
(type III) 2HDM in
which the components of the second scalar doublet
have large FCNC couplings to top quarks, but only a small mixing with the
SM-like doublet. Working in a basis in which only one of the two scalar
doublets $\Phi_1$, $\Phi_2$ has
a non-zero vacuum expectation value (vev), we decompose $\Phi_1$ and $\Phi_2$
into
\begin{align}
  \Phi_1 = \begin{pmatrix}
             G^+ \\
             \frac{1}{\sqrt{2}} (v + h_1 + i G^0)
           \end{pmatrix}
  \qquad
  \Phi_2 = \begin{pmatrix}
             H^+ \\
             \frac{1}{\sqrt{2}} (h_2 + i h_3)
           \end{pmatrix} \,.
  \label{eq:Phi1-Phi2}
\end{align}
Here $G^\pm$ and $G^0$ are the Goldstone bosons that are eaten by the $W^\pm$
and the $Z$. In the absence of CP violation, $h_1$ and $h_2$ mix into the CP even
physical scalars $h$ and $H^0$, and $h_3$ is identified with the CP odd scalar $A^0$.
The scalar potential $V$ of the model depends on three dimensionful
parameters and seven dimensionless couplings.  The condition that $\Phi_1
= (0, v/\sqrt{2})$, $\Phi_2 = (0, 0)$ should be a minimum of $V$ eliminates
two of these parameters, and one further coupling can be dropped here because
it is only relevant for scalar self-interactions. The seven remaining parameters
can be expressed in terms of the masses $m_h$, $m_{H^0}$, $m_{A^0}$, $m_{H^\pm}$,
the $h$--$H^0$ mixing $\sin\alpha$, and two dimensionless parameters $\lambda_3$
and $\lambda_7$.~\cite{Buschmann:2016uzg}  The Yukawa couplings of the two scalar
doublets are given in the up quark sector by
\begin{align}
  \mathcal{L}_\text{up}
    &= - \eta_1^{ij} \overline{Q_L^i} \tilde\Phi_1 u_R^j
       - \eta_2^{ij} \overline{Q_L^i} \tilde\Phi_2 u_R^j
       + h.c. \,,
  \label{eq:L-up}
\end{align}
with $\tilde\Phi_k \equiv i \sigma^2 \Phi_k^\dag$.  The couplings to down quarks
and leptons are analogous.

\begin{figure}
  \begin{center}
    \begin{tabular}{cc}
      \includegraphics[width=0.48\textwidth]{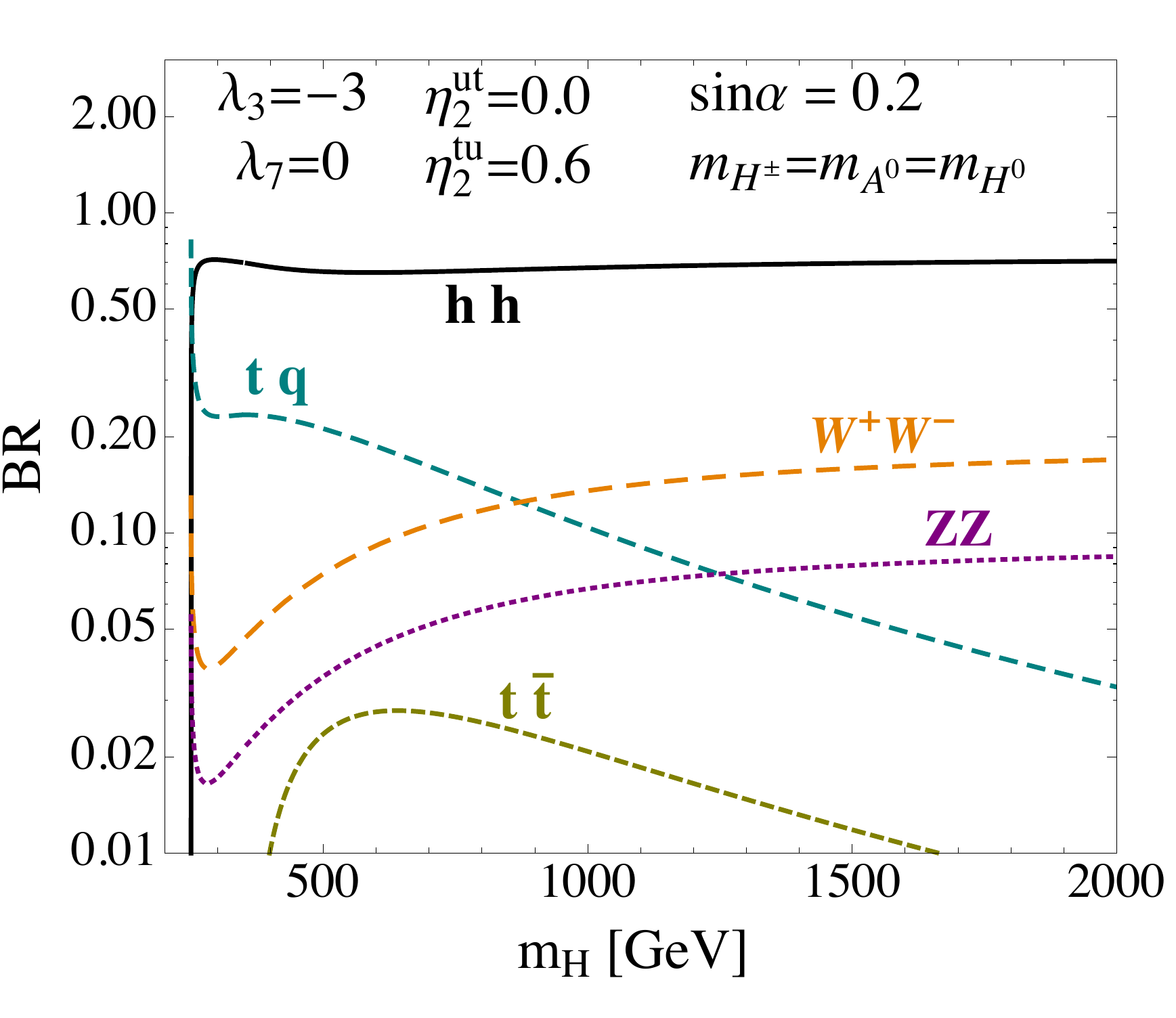} &
      \includegraphics[width=0.50\textwidth]{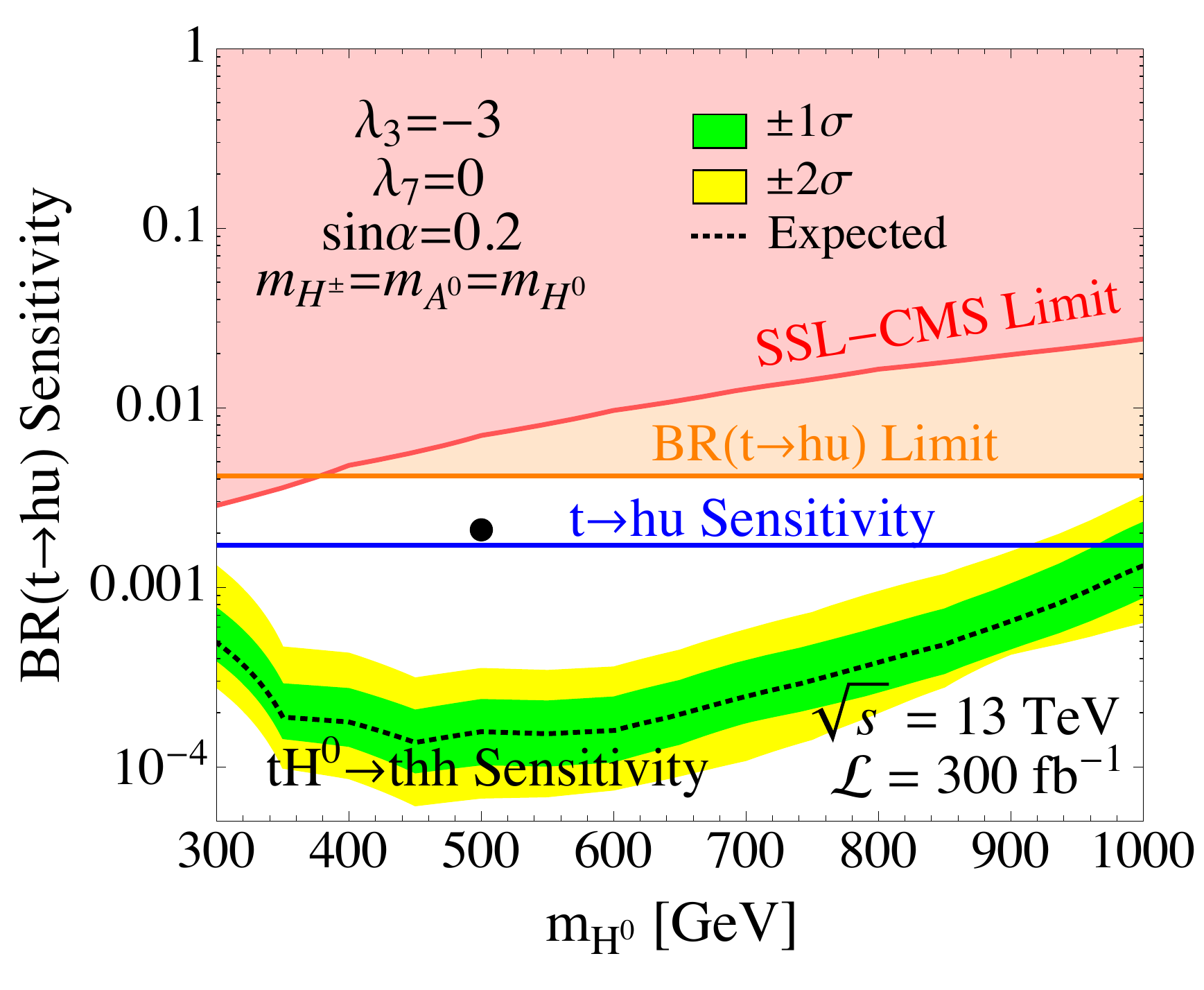} \\
      (a) & (b)
    \end{tabular}
  \end{center}
  \caption{(a) Branching ratios of the $H^0$ boson for one parameter point in a
    quark flavor violating 2HDM.~\protect\cite{Buschmann:2016uzg}
    (b) Sensitivity to flavor changing couplings
    of the top quark to the $h$ and $H^0$ bosons at the same parameter point,
    expressed here in terms of the branching ratio for $t \to h q$. The orange
    and blue lines show the current and conservative future sensitivities,
    respectively, to $t \to h u$ using multi-lepton and lepton + di-photon
    final states~\protect\cite{Greljo:2014dka}. The red shaded region is the
    limit from a CMS search for same-sign di-leptons + $b$
    jets~\protect\cite{Chatrchyan:2012paa}.  The Brazilian band shows the predicted
    sensitivity for the search discussed in the text.~\protect\cite{Buschmann:2016uzg}}
  \label{fig:thh}
\end{figure}

In a 2HDM, flavor changing effects become much more accessible once not only
the lightest scalar $h$, but also its heavier companions can be directly
produced.  An example is the process $p p \to t H^0$, mediated by a flavor
violating $t u H^0$ coupling and followed by the decay $H^0 \to h
h$.~\cite{Buschmann:2016uzg} If the (flavor conserving and flavor violating)
couplings of $H^0$ to quarks are not too large, this decay mode can be dominant
in large regions of parameter space, as illustrated in \cref{fig:thh} (a) for a particular
parameter point of the quark flavor violating 2HDM.  The most favorable final state to look
for $p p \to t + (H^0 \to h h)$ consist of one lepton (from top decay), five
$b$ quarks, and missing transverse energy.  It can be extracted from the
background by requiring one isolated, positively charged lepton, at least five
jets, and at least four $b$ tags.~\cite{Buschmann:2016uzg} Moreover, $p_T$ and
$\eta$ cuts need to be imposed on these objects, as well as on the
reconstructed $h$ bosons.  Combinatorial backgrounds can be suppressed by
optimizing the invariant masses of two jet pairs and of the fifth jet, the
lepton, and the missing energy.

The sensitivity of such a search is illustrated in \cref{fig:thh} (b). The parameter
point chosen for this plot was selected such that the $tuH^0$ coupling is the
dominant Yukawa coupling of $H^0$, i.e.\ that $\eta_2^{tu}$ is the only relevant
entry of the matrix $\eta_2$ from \cref{eq:L-up}. (Note that the complementary
coupling $\eta_2^{ut}$ cannot be too large because of constraints from $B$ meson
mixing.~\cite{Buschmann:2016uzg})  We see that the proposed search for
$p p \to t + (H^0 \to h h)$ is very promising and can easily supersede current
and future limits from other search channels.

\section{New Probes of FCNC Couplings to Leptons}

\begin{figure}
  \begin{center}
    \begin{tabular}{cc}
      \includegraphics[width=0.48\textwidth]{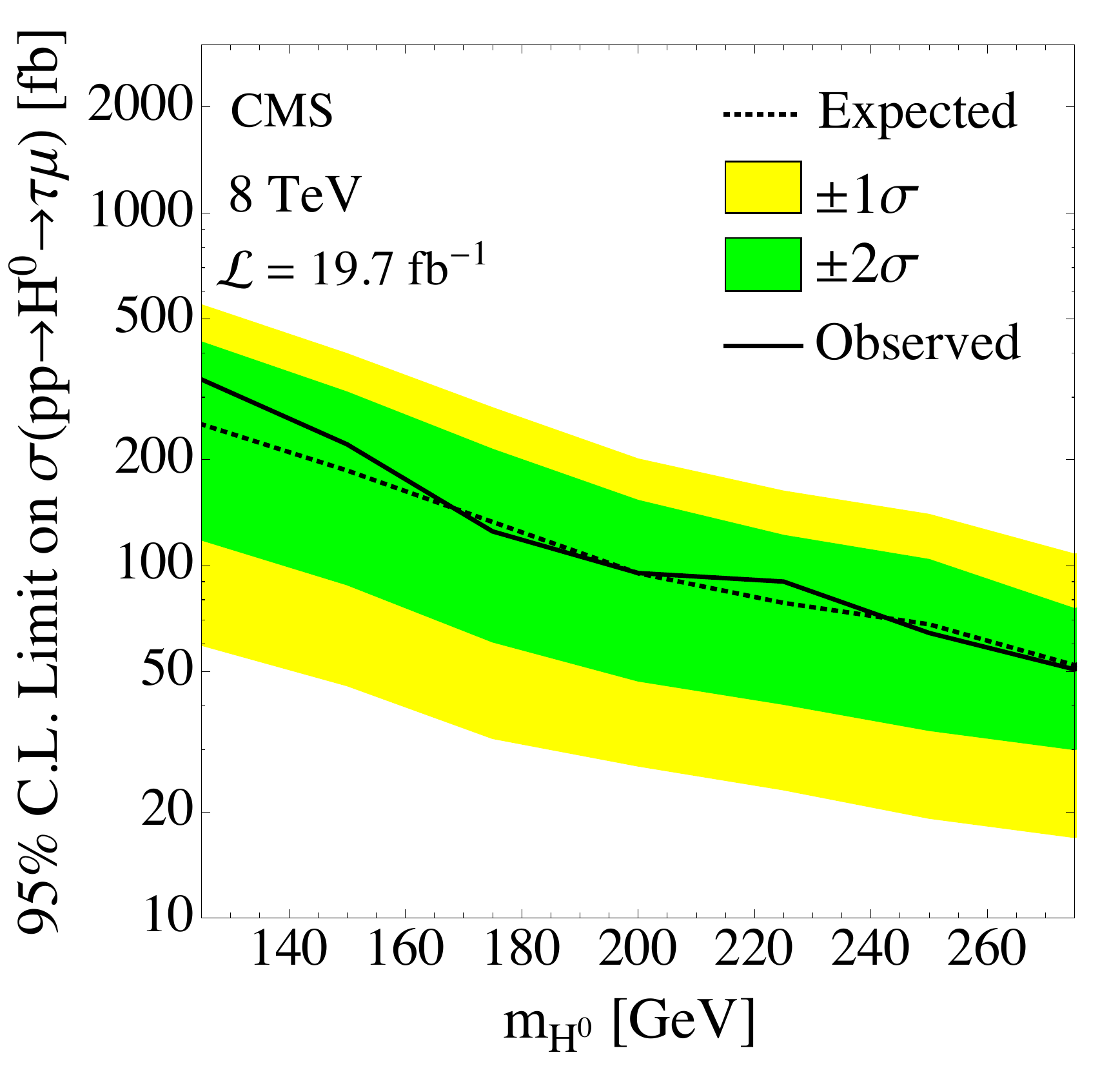} &
      \includegraphics[width=0.48\textwidth]{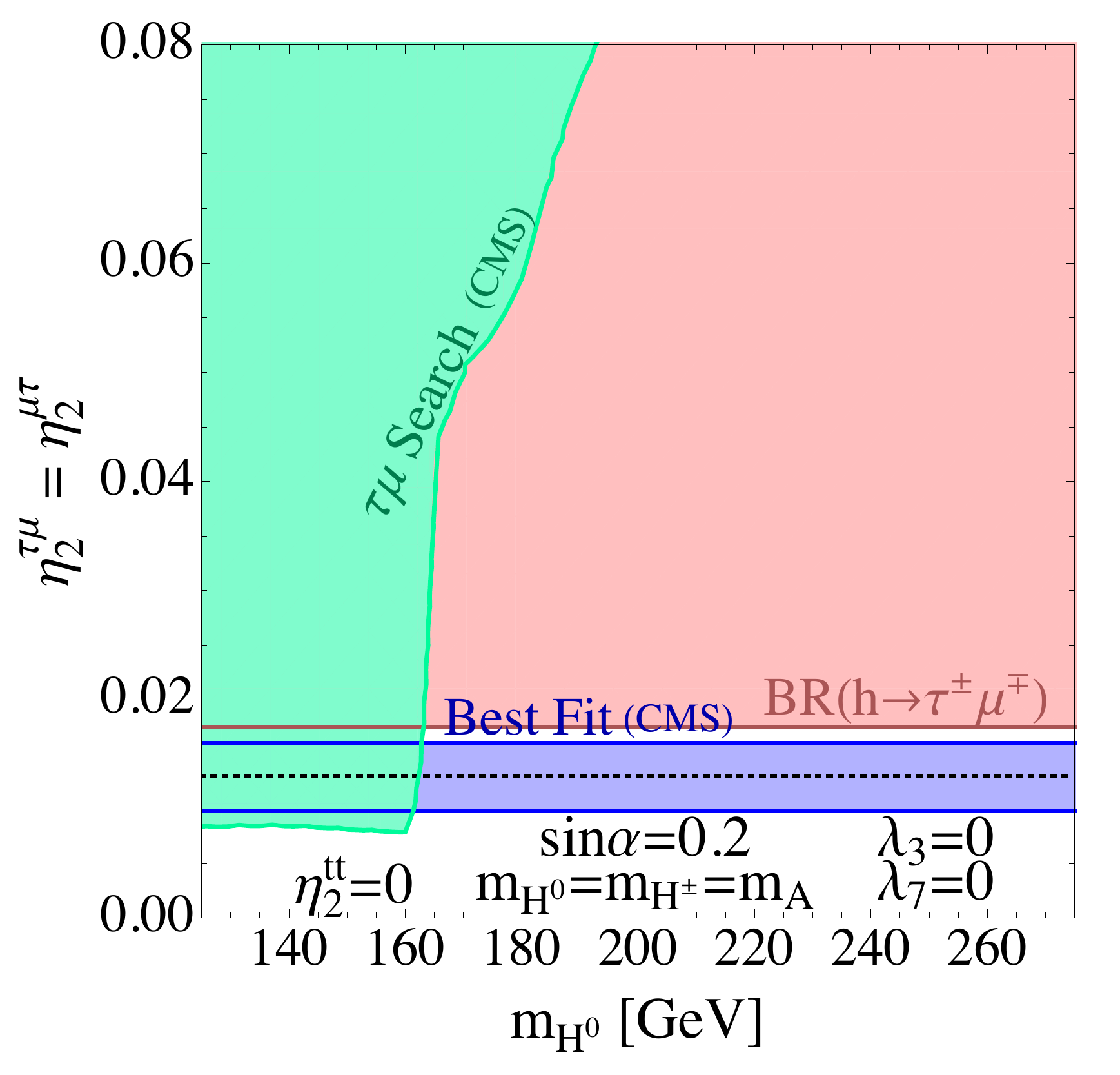} \\
      (a) & (b)
    \end{tabular}
  \end{center}
  \caption{(a) Constraints on the FCNC decay $H^0 \to \tau\mu$ of the heavy CP even
    scalar $H^0$ in a 2HDM, derived by recasting the CMS search for
    $h \to \tau\mu$~\protect\cite{Khachatryan:2015kon,Buschmann:2016uzg}. (b) Resulting
    constraints on the FCNC Yukawa coupling $\eta_{\tau\mu}$ and $\eta_{\mu\tau}$
    of the second scalar doublet (green) compared to the limit from the
    direct search for $h \to \tau\mu$ (red)~\protect\cite{Khachatryan:2015kon} and
    the region favored by the small excess in that search (blue $1\sigma$
    band).}
  \label{fig:Htaumu}
\end{figure}

Let us now turn to flavor violating couplings of the scalar sector to leptons.
Also in this case, the sensitivity in specific model frameworks can be
much larger than for the simple EFT from \cref{eq:L-eft}.  We consider
again a type III 2HDM and investigate to what extent current searches for $\mu$--$\tau$
resonances~\cite{Khachatryan:2015kon} constrain the FCNC decay $H^0 \to \tau\mu$.
As for FCNC in the quark sector, this decay is expected to have a much large branching
ratio than $h \to \tau\mu$, which offsets the smaller production cross section of $H^0$.
Recasting the CMS search for
$h \to \tau\mu$~\cite{Khachatryan:2015kon,Buschmann:2016uzg}, we find the limits shown
in \cref{fig:Htaumu}.  Note that above $m_{H^0} = 2 m_W$, the dominant
$H^0$ decay mode becomes $H^0 \to W^+ W^-$, reducing the branching ratio
to the $\tau\mu$ final state and thus limiting the parameter sensitivity in this
mass range, see \cref{fig:Htaumu} (b).

\section{CP Violation in FCNC $h$ Decays}

\begin{figure}
  \begin{center}
    \includegraphics[width=\textwidth]{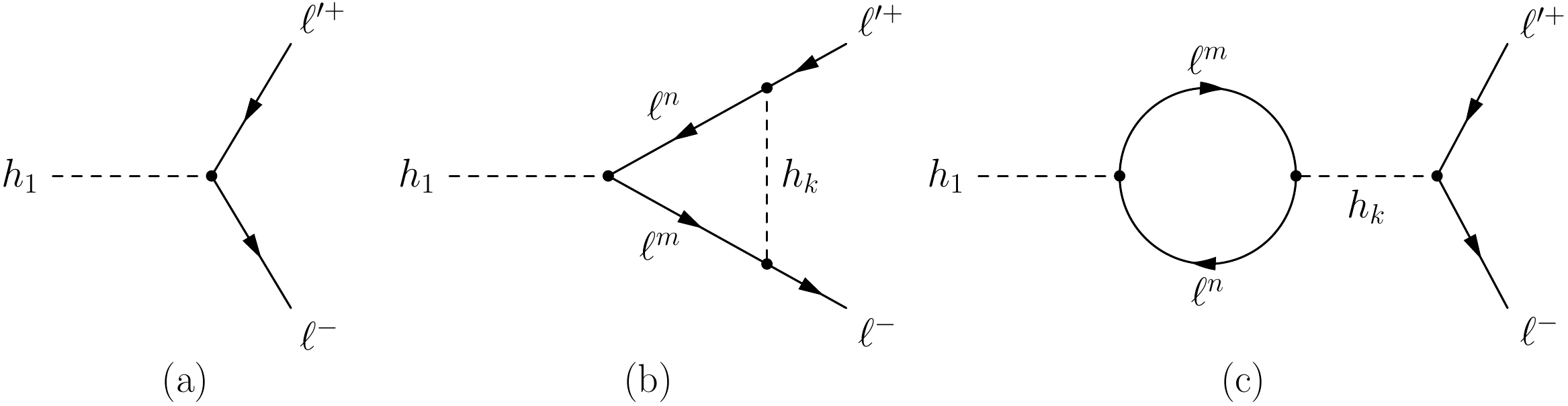}
  \end{center}
  \caption{The tree level and 1-loop diagrams in a 2HDM whose interference can lead
    to a CP violating asymmetry between $h \to \tau^+ \mu^-$ and $h \to \tau^-
    \mu^+$ final states. In these diagrams, $h_k$ ($k=1,2,3$) are the three
    neutral scalars and $\ell$, $\ell'$, $\ell^m$ are charged leptons.}
  \label{fig:cpv-higgs}
\end{figure}

If flavor violation in the scalar sector is discovered---for instance if the
CMS hint
for $h \to \tau\mu$~\cite{Khachatryan:2015kon} should be corroborated with 13~TeV data---it
stands to reason to ask whether it could be accompanied by CP violation.  In
fact, the resulting signature---an asymmetry between $h \to \tau^+ \mu^-$ and
$h \to \tau^- \mu^+$ would offer a very direct probe of CP violation.
\Cref{fig:cpv-higgs} illustrates how such an asymmetry
could arise in a type III 2HDM from the interference between tree level and 1-loop
processes.  A detailed phenomenological analysis~\cite{Kopp:2014rva} shows
that, like all searches for CP violation in the scalar sector, also the search for
CP violation in $h \to \tau\mu$ requires very large integrated luminosity before
one can hope to observe a signal. The reason is that a loop-suppressed effect
must be observed on top of an already small branching ratio.  Assuming
an $h \to \tau\mu$ signal at the level hinted at by CMS~\cite{Khachatryan:2015kon},
a detection of a CP asymmetry may be possible at the high-luminosity LHC if the
mixing angle between the SM-like $h$ boson and its heavy partners is small
and if the heavy scalars are close to each other in mass.~\cite{Kopp:2014rva}

\section{Summary}

In summary, we have reviewed from a phenomenologist's point of view the current
status of FCNC searches in the scalar sector.  We have outlined a number of possible
directions for future experimental work, including in particular
(i) searching explicitly for the so-far neglected process $p p \to t h$, which
could be exploited to distinguish $tuh$ and $tch$ couplings in the event of a discovery;
(ii) exploiting new final states, for instance in the
fully hadronic processes
$p p \to (t \to W b) + (t \to h q) \to \text{hadrons}$ and $p p \to t h \to \text{hadrons}$;
(iii) searching for
FCNC decays of heavy scalar bosons, which can have very large branching ratios
in 2HDMs;
(iv) Searching for CP violation in $h \to \tau\mu$.

\section*{Acknowledgments}

It is a great pleasure to thank the organizers and participants of the 51st
Rencontres de Moriond for an exciting conference, great talks, and interesting
discussions in the lecture hall and on the ski slopes.  The work presented
in this talk is based on several papers written in collaboration with
Malte Buschmann, Admir Greljo, Roni Harnik, Jernej Kamenik, Jia Liu,
Marco Nardecchia, Xiao-Ping Wang, and Jure Zupan. Thank you for the great
collaborations we had and will hopefully have in the future!
Parts of this work were supported by Fermilab, by the Max Planck Society,
by the German Research Foundation (grants FOR~2239 and KO~4820/1--1),
by the German Federal Ministry for Education and Research (PRISMA Cluster of
Excellence) and by the European Research Council (grant No.\ 637506).

\bibliographystyle{kpmoriond}
\bibliography{higgs-fv}

\end{document}